\documentstyle[11pt]{article}
\def\th{\theta}
\def\ha{\hat a}\def\hb{\hat b}\def\hc{\hat c}\def\hd{\hat d}
\def\he{\hat e}\def\hf{\hat f}
\def\k{\kappa}
\def\del{\partial}
\def\d{\delta}
\def\e{\epsilon}
\def\h{\eta}

\let\la=\label

\let\bm=\bibitem

\def\nn{\nonumber}
\def\bd{\begin{document}}
\def\ed{\end{document}}
\def\be{\begin{equation}}
\def\ee{\end{equation}}
\def\ba{\begin{array}}
\def\ea{\end{array}}
\def\bea{\begin{eqnarray}}
\def\eea{\end{eqnarray}}
\def\fft#1#2{{#1 \over #2}}
\def\sst#1{{\scriptscriptstyle #1}}
\def\ft#1#2{{\textstyle{\sst#1\over\sst #2}}}
\def\det{{\rm det\,}}
\def\tr{{\rm tr}}
\def\hm{{\hat m}}
\newcommand{\eq}[1]{(\ref{#1})}
\newcommand{\ho}[1]{$\, ^{#1}$}
\newcommand{\hoch}[1]{$\, ^{#1}$}
\newcommand{\tamphys}{\it\small Center for Theoretical Physics,
Texas A\&M University, College Station, TX 77843, USA}
\newcommand{\newton}{\it\small Isaac Newton Institute for Mathematical Sciences,
Cambridge, UK}
\newcommand{\kings}{\it\small Department of Mathematics, King's College,
London, UK}

\newcommand{\auth}{\large P.S. Howe\hoch{1}, E. Sezgin \hoch{2\dagger}
            and P.C. West \hoch{3\ddagger}}

\begin{document}

\hfill{KCL-TH-97-11}

\hfill{CTP TAMU-12/97}

\hfill{NI-97008}

\hfill{hep-th/9702111}

\hfill{\today}

\vspace{20pt}

\begin{center}
{\Large The Six-Dimensional Self-Dual Tensor }

\vspace{30pt}

\auth

\vspace{15pt}

\begin{itemize}
\item [$^1$] \kings
\item [$^2$] \tamphys
\item [$^3$] \newton
\end{itemize}

\vspace{60pt}

\end{center}

{\bf Abstract}

The equations of motion for a self-interacting self-dual tensor in six
dimensions are extracted from the equations describing the $M$-theory
five-brane. These equations are presented in a self-contained,
six-dimensional Lorentz-covariant form. In particular, it is shown that
the field-strength tensor satisfies a non-linear generalised
self-duality constraint. The self-duality equation is rewritten in
five-dimensional notation and shown to be identical to the corresponding
equation in the non-covariant formalism.

{\vfill\leftline{}\vfill
\vskip	10pt
\footnoterule
{\footnotesize
\hoch{\dagger} Research supported in part by NSF Grant	PHY-9411543 \vskip
-12pt}
\vskip	10pt
{\footnotesize
\hoch{\ddagger} Permanent Address: \kings \vskip -12pt}}

\pagebreak
\setcounter{page}{1}

\section{Introduction}

Super $p$-branes play a central r\^{o}le in string duality and in
$M$-theory, and it is therefore important to understand their
properties. One aspect of super $p$-branes which can be studied with
available mathematical tools is the construction of the worldsurface
actions that describes their dynamics. A well-known property of all the
existing super $p$-brane actions is that in addition to being
worldsurface reparametrization invariant, they possess a fermionic gauge
symmetry, called $\k$-symmetry. Gauge-fixing of these symmetries leads
to worldsurface supersymmetric field theories that describe highly
nonlinear self-interactions of matter supermultiplets.

The focus of this paper will be on the $M$-theory five-brane for which
the resulting worldsurface multiplet contains a chiral two-form. The
full equations of motion for this object were given in \cite{hs2}, using
a concise superspace language. More recently, full actions for the same
object have been constructed in \cite{f8,f9} in different approaches,
and the component version of the results given in \cite{hs2} has been
obtained \cite{hsw1}. An interesting aspect of these results is the
manner in which the self-duality equation for the interacting chiral
two-form arises. In an earlier work that led to \cite{f8}, Perry and
Schwarz \cite{ps} constructed a self-duality equation with manifest
five-dimensional covariance, and they suggested that it was impossible
to make manifest the hidden six-dimensional covariance. On the other
hand, the superspace formalism of \cite{hs2} and the explicit component
results of \cite{hsw1} involved just such a system.

In view of these developments, it is of some interest to study the
self-dual tensor in six dimensions by itself, i.e. extracted from the
brane context. This is what we shall do in this paper. We begin in the
next section by extracting these equations in terms of a self-dual
three-form field of the type introduced in \cite{hs1} and then show that
there is a new three-form which satisfies the Bianchi identity and a
generalised, but still manifestly six-dimensional covariant, duality
constraint. In the following section we rewrite our equations with only
manifest five-dimensional covariance in order to study the relationship
between our results \cite{hs2, hsw1} and those of \cite{ps}. The result
we find is that our six-dimensional covariant self-duality equation,
when expressed in terms of five-dimensional fields, reduces precisely to
that of Perry and Schwarz. Section 4 contains some concluding remarks.

\section{The 6D Covariant Self-Duality Equation}

In the superspace approach to the five-brane in eleven dimensions the
antisymmetric tensor makes its first appearance as a self-dual tensor
$h_{abc}$ which occurs directly in the embedding. This field (or rather
its leading component in a $\th$ expansion) is not directly related to a
two-form potential, but it can be shown that there is superspace
three-form $H_3$ which is; in fact this three-form satisfies
\be
dH_3=-\ft1{4}H_4\ ,
\la{bi}
\ee
where $H_4$ is the pull-back of the four-form of eleven-dimensional
supergravity. The purely vectorial component of this three-form, which
we denote by $H'_{abc}$, is related to $h$ by
\be
H'_{abc}=m_a{}^d m_b{}^e h_{cde} \la{2}
\ee
where
\be
m_a{}^b := \d_a^b-2 k_a{}^b\ , \la{m}
\ee
and where
\be
k_a{}^b := h_{acd}\,h^{bcd}\ .\la{k}
\ee

We shall extract from the brane equations the equations describing a
self-interacting self-dual tensor field in six-dimensional flat
spacetime by setting all the other fields equal to their flat space
values. The equations are self-duality,
\be
h^{abc}={1\over 3!}\e^{abcdef}\,h_{def}\ . \la{sd1}
\ee
with $\e^{012345}=+1$ and $\h_{ab}={\rm diag}(-1,+1,\ldots ,+1)$, and
the equation of motion for $h$ which becomes
\be
m^{ab}\del_a h_{bcd}=0\ . \la{eom}
\ee
Equations \eq{sd1} and \eq{eom} are equations in ordinary flat
six-dimensional spacetime as are all subsequent equations in this paper.
The goal is to rewrite these equations in terms of the more familiar $H$
field. In fact equation \eq{2} has been written in a slightly unusual
basis and it is necessary to correct this in order to find the relation
between the components of $H$ in a coordinate basis and the components
of $h$. This turns out to be
\be
H_{abc}= (m^{-1})_a{}^d\,h_{bcd}\ . \la{hh}
\ee
Through this relation the self duality of $h_{bcd}$ imposes a self
duality condition on $H_{abc}$ which, as we will show below,
is a rather  complicated condition.

Since these equations were derived from the superspace system the
Bianchi identity \eq{bi} will ensure that $dH=0$ in the truncated
theory, where $H$ is now the spacetime three-form. However, to be
completely self-contained we shall show that the Bianchi identity for
$H$ can be derived from the basic equations for $h$. It follows from
self-duality that the tensor $k$ introduced in \eq{k} is traceless and
that its square is proportional to the unit tensor, i.e., in matrix
notation,
\be
k^2= \ft16 \tr\,k^2\ , \la{ti}
\ee
and that
\be
h_{abe}h^{cde}= -\d_{[a}{}^{[c}\,k_{b]}{}^{d]}\ , \la{hhk}
\ee
Therefore
\be
m^{-1}=Q^{-1}(1+2k)=Q^{-1}(2-m)\ , \la{inv}
\ee
where we have introduced
\be
Q=1-\ft2{3}\tr\,k^2\ .
\ee
We note in passing that the above expession for $m^{-1}$ shows that $H$
defined by \eq{hh} is indeed totally antisymmetric since duality
implies that $k_a{}^d h_{bcd}$ is totally antisymmetric and
anti-self-dual (while $k_a{}^d k_b{}^e h_{cde}$ is totally antisymmetric
and self-dual). Using this expression for $m^{-1}$ we find
\bea
\e_{abcdef}\del^cH^{def}&=&6\del^c(Q^{-1}m_c{}^d h_{abd})\\
                        &=&6\del^c \left(Q^{-1}m_c{}^d\right)h_{abd}\ ,
\eea
where in the second step we have used the equation of motion \eq{eom}.
It is not difficult to show, again using the equations of motion, that
\be
m^{ab}\del_am_{bc}=0\ .
\ee
Using \eq{inv} in this equation we find
\be
m^{ab}\del_a(Q(m^{-1})_{bc})=0\ ,
\ee
and this leads, after a little algebra, to
\be
\del^a(Q^{-1}m_{ab})=0\ .
\ee
Hence we have established
\be
\e^{abcdef}\del_c H_{def}=0\ ,
\ee
as required.

We next translate the self-duality condition on $h$ into a generalised
self-duality condition for $H$. Splitting $H$ into its self-dual and
anti-self-dual parts, $H^+$ and $H^-$, one finds from \eq{hh} that
\bea
H^+_{abc}&=&Q^{-1}h_{abc}\ ,\\
H^-_{abc}&=&Q^{-1}k_a{}^d h_{dbc}\ .
\eea
If we define
\be
K_a{}^b=H^+_{acd} H^{+bcd}\ ,
\ee
then
\be
K_{ab}=Q^{-2} k_{ab}\ .
\ee
Now
\be
H^+\cdot H^-:=H^+_{abc} H^{-abc}=Q^{-2}\tr k^2=Q^2 \tr K^2\ ,
\ee
and
\be
H^- = Q^2 K_a{}^d H^+_{bcd}\ ,
\ee
so that we finally derive
\be
H^- = (\tr K^2)^{-1} (H^+\cdot H^-) K_a{}^d H^+_{bcd}\ ,
\la{sd2}
\ee
where we recall that $K_a{}^d H^+_{bcd}$ is totally antisymmetric and
anti-self-dual as a consequence of the self-duality of $H^+$.
This is the self-duality condition we were looking for. At first sight
it does not appear to determine $H^-$ in terms of $H^+$ since it is
clearly invariant under rescalings of $H^-$. However, the equation has
been derived by purely algebraic manipulations and does not take into
account the Bianchi identity which must also be satisfied. When one does
this one finds that the freedom to rescale $H^-$ disappears.

\section{The Self-dual Tensor in 5D Notation}

In this section we will use hatted indices for six dimensions,
and unhatted ones for five dimensions. For clarity we shall also put a hat on the six-dimensional $m$-matrix. Our main purpose is to
analyse \eq{hh} in five-dimensional language. We begin by defining
\bea
f_{ab} &:=& h_{ab5}\ , \la{fh1}\\
F_{ab} &:=& H_{ab5}\ . \la{fh2}
\eea
Hence
\be
h_{abc}=\ft12 \e_{abcde}\,f^{de}\ . \la{h3}
\ee
Inverting \eq{hh} we get
\be
h_{\ha\hb\hc}=\hm_{\ha}{}^{\hd} H_{\hb\hc\hd}
\la{hh1}
\ee
from which it follows that
\be
f_{ab}=m_a{}^c\,F_{cb}\ . \la{ff}
\ee
Finally, let us define
\be
{\tilde H_{ab}}=\ft1{3!} \e_{abcde}\,H^{cde}\ . \la{ht}
\ee
Our goal is now to express ${\tilde H}$ in terms of $H_{ab5}:= F_{ab}$.
To this end, we begin by expressing the matrix $m$ and its inverse in
five-dimensional notation. Recalling the definition \eq{m}, one finds
\bea
m_a{}^b   &=& \d_a{}^b(1-2\tr\,f^2)+ 8(f^2)_a{}^b\ , \la{mab}\\ 
m_a{}^5   &=& -\e_{abcde}\,f^{bc}\,f^{de}\ , \la{ma5}\\
m_5{}^5   &=& (1+2\tr\,f^2)\ \la{m55} .
\eea
The components of the inverse matrix  can be calculated from \eq{inv}:
\bea
(m^{-1})_a{}^b  &=& Q^{-1}\left[\d_a{}^b(1+2\tr\,f^2)-
                    8(f^2)_a{}^b\right]\ , \\
(m^{-1})_a{}^5  &=& Q^{-1}\e_{abcde}\,f^{bc}\,f^{de}\ ,\\
(m^{-1})_5{}^5  &=& Q^{-1}(1-2\tr\,f^2)\ , \la{mi2}
\eea
where
\be
Q=1+ 4(\tr\,f^2)^2-16\tr\,f^4\ . \la{q2}
\ee
From \eq{hh} we have
\be
H_{abc}= (m^{-1})_a{}^d\,h_{dbc} + (m^{-1})_a{}^5\,f_{bc}\ .
\ee
Taking the dual of  this equation, and using the definitions and
formulae above,
one finds
\be
{\tilde H}_{ab} = -Q^{-1}\,m_a{}^c\,f_{cb}\ , \nonumber\\ \la{ht1}
\ee
or, in matrix notation,
\be
{\tilde H} = -Q^{-1}\left[(1-2\tr\,f^2)f + 8f^3\right]\ .
\la{ht2}
\ee

It is straightforward even at this stage to write the matrix part of the
above equation in terms of $F$. Using the relation $f=mF$, the identity
$F_a{}^c m_c{}^5=0$, which again follows directly from equation \eq{hh}
and the above expressions for $m$, we can write 
\be
{\tilde H}_{ab} = - Q^{-1}F_a{}^c G_{cb}\ , \la{hgf}
\ee
where
\bea
G_{ab} &:=& m_a{}^{\hat c} m_{\hat c b} \la{gmm}\\
       &=& (2 -4\tr f^2 -Q)\h_{ab} + 16(f^2)_{ab}\ , \la{g} 
\eea
which follows from \eq{mab}, \eq{ma5} and \eq{q2}. Multiplying \eq{gmm}
with $F^2$, using $F_a{}^c m_c{}^5=0$ and recalling that $f=mF$, we
derive $GF^2=f^2$. Using this relation in \eq{g} we can solve for $G$.
Substituting the result in \eq{hgf} we find
\be
{\tilde H} = - Q^{-1}(2 -4\tr f^2 -Q) \left(F\over 1-16 F^2\right)\ . 
\ee

To find the final result we must express the traces of $f$ in terms of
$F$. At this point, it is important to recall a well known identity that
holds for {\it any} antisymmetric $5\times 5$ matrix $X$:
\be
X^5 = \ft14 \left[ \tr\,X^4 - \ft12(\tr\,X^2)^2\right]\,X +
      \ft12 (\tr\,X^2)\,X^3\ . \la{x5}
\ee
Applying this identity to the matrix $F$, we write
\be
F^5 = (y_2-\ft12 y_1^2)\,F + y_1\,F^3\ , \la{f5}
\ee
where
\bea
y_1 &=& \ft12 \tr\,F^2\ , \nn\\
y_2 &=& \ft14 \tr\,F^4\ .  \la{y12}
\eea

The next step in the analysis of \eq{ht2} is to express $f$ in terms of
$F$. From the main duality equation \eq{ff}, and the general identity
\eq{x5}, it follows that $f$ necessarily has the form
\be
f= a\,F + b\,F^3\ ,\la{f12}
\ee
where $a$ and $b$ are functions of $F$ that can be determined from
the duality equation \eq{ff}, reproduced here for the reader's convenience:
\be
f=(1-2\tr\,f^2 + 8f^2)\,F\ . \la{ff2}
\ee
The $f^2$ term can easily be computed from \eq{f12} and \eq{f5}:
\be
f^2=\left[a^2+ b^2 (y_2-\ft12 y_1^2)\right]\,F^2 +
\left(2ab+b^2 y_1 \right)\,F^4\ . \la{f2}
\ee
All the terms in \eq{ff2} can be computed in a similar manner. At the
end, comparing the coefficients of the terms proportional to $F$ and
$F^3$, one finds
\bea
&& a +\ft12 b y_1 = 1 \ , \la{e1}\\
&& a^2 + 2  ab y_1 +  b^2 (y_2+\ft12 y_1^2)=\ft18 b\ .\la{e2}
\eea
The solution of these equations is given by
\be
a_\pm = {32-y_1\pm y_1 Z\over 32y_2-8y_1^2}\ ,\quad\quad\quad
b_\pm = {1-8y_1\pm Z\over 16y_2-4y_1^2}\ , \la{sol}
\ee
where
\be
Z = \sqrt {1-16y_1+128 y_1^2 -256 y_2} \la{z}
\ee
The equations \eq{e1} and \eq{e2} will be used frequently in the
calculations to simplify the resulting expressions. It turns out that we
will not need to use \eq{sol} and therefore the final result is independent 
of the sign ambiguity in this equation.

Having determined $f$ explicitly in terms of $F$, we next evaluate
$\tr\,f^2$, $f^3$ and $Q$, in order to express \eq{ht2} in terms of $F$
alone. An expression for $\tr\,f^2$ is easily obtained from \eq{f2}. The
calculation of $f^3$ requires a bit of work, but is straighforward. With
the aid of \eq{e1} and \eq{e2}, we find
\be
f^3 = \left( 2y_2-y_1^2+\ft18 b \right)\,F
      +\ft18 \left[ a+ b (y_1-y_1^2+2y_2)\right]\,F^3\ .\la{f3}
\ee

Putting together the results obtained so far, we find that the self-duality
equation
\eq{ht2} takes the simple form
\be
{\tilde H}= -\ft18 b Q^{-1} \left[ (1-16y_1)F+16F^3\right]\ . \la{th2}
\ee

There remains the evaluation of $Q$. Starting from \eq{q2}, and using
\eq{f2} to calculate $(\tr\,f^2)^2$ and $\tr\,f^4$, we find, with the
aid of \eq{e1} and \eq{e2}, the result
\be
Q=\ft18 b Z\ .
\ee
Therefore, recalling the definition of $Z$, and rescaling $F\rightarrow
\fft14 F $, $H\rightarrow -\fft14 H$, we deduce the final result
\be
{\tilde H} = {(1-y_1)F+F^3\over \sqrt{1-y_1+\ft12 y_1^2-y_2}}\
= \sqrt{1-y_1+\ft12 y_1^2-y_2}\left({ F\over 1-F^2}\right)\ . \la{fr}
\ee
This formula is precisely the self-duality equation of Perry and Schwarz
\cite{ps}
(their equation (51)).

\section{Conclusions}

Our main result is that there does indeed exist a self-interacting
self-dual tensor in six dimensions for which the generalised duality
relation and the equations of motion can be written in manifestly
six-dimensional covariant form. Furthermore, the self-duality equation
which arises naturally in the description of the M-theory five-brane
reduces precisely to the self-duality equation of Perry and Schwarz,
when expressed in five-dimensional language. Starting from the
Perry-Schwarz version, it may not have been obvious in the past how to
obtain manifest six-dimensional covariance. However, we now know the
answer: first, one combines $H_{abc}=-\ft12\e_{abcde}\tilde H^{de}$ and
$F_{ab}$ into a six-dimensional field strength as
\be
(H_{abc}, F_{ab}) \rightarrow H_{\ha\hb\hc}\ ,
\ee
where $\ha$ is now a six dimensional index, $\ha=(a,5)$, and $F_{ab} :=
H_{ab5}$. Next, one defines three-form $h_{\ha\hb\hc}$
by
\be
h_{\ha\hb\hc}=\left(\d_{\ha}^{\hd}-2h_{\ha\he\hf}\,h^{\hd\he\hf}\right)\,
H_{\hb\hc\hd}\ , \la{6d}
\ee
then, as a consequence of the five-dimensional duality relation,
the derived field $h_{\ha\hb\hc}$ is self-dual.

It is interesting that this equation arises naturally in the supersurface
embedding approach to the formulation of super $p$-branes. The approach
is by no means new, and it has been applied to various $p$-branes in the
past, with varying degrees of success. We refer the reader to \cite{hs1}
for a brief review of the extensive literature on the subject. The point
we wish to emphasize here is that the approach is very natural, and its
power becomes apparent when one gets to terms with the idea of searching
for the equations of motion first and later deriving an action, when possible.

The beauty of the superembedding approach is that not only it treats
the worldsurface and target space supersymmetry on an equal footing by
considering the embedding of a world supersurface into the target
superspace, but it applies to {\it all} super $p$-branes, regardless of the
precise nature of the worldsurface multiplet, in a universal way that
has a natural geometrical interpretation. Indeed, new kinds of super
$p$-branes, e.g. $L$-branes, which have matter supermultiplets other
than those considered so far have been suggested in \cite{hs1}, and we
have already preliminary results which show that things work in much the
same way they do in the case of M-theory five-brane. We should expect,
therefore, the uncovering of interesting new results on nonlinear
self-couplings of matter supermultiplets in diverse dimensions which
have not been explored so far.

Finally, it would be interesting to
investigate various aspects of the $M$-theory five-brane self-duality
equation discussed in this paper equation further. We have already exhibited 
the six-dimensional self-dual tensor equations as part of the full
set of equations for the five-brane in an eleven-dimensional supergravity background \cite{hs2,hsw1}. A lot remains
to be done, however, and we will report elsewhere \cite{hsw2} on further
aspects of the emerging picture of the $M$-theory
five-brane which seems to possess many interesting features.

{\bf Acknowledgements.}
We thank Michael Duff and Per Sundell for many stimulating discussions
on super five-branes.

\pagebreak

\end{document}